\documentclass[amsmath,amssymb,pra]{revtex4}
\usepackage[latin1]{inputenc}
\usepackage{graphicx}% Include figure files
\usepackage{dcolumn}% Align table columns on decimal point
\usepackage{bm}% bold math
\usepackage{color}

\begin{document}
\title{Repeaters in relativistic communications}
\author{Juan Carlos Garc\'ia-Escart\'in}
\email{juagar@tel.uva.es}
\author{Pedro Chamorro-Posada}
\affiliation{Dpto. de Teor\'ia de la Se\~{n}al y Comunicaciones. ETSI de Telecomunicaci\'on. Universidad de Valladolid. Campus Miguel Delibes. Paseo Bel\'en 15. 47011 Valladolid. Spain.}
\date{\today}
\begin{abstract}
The communication efficiency between a transmitter and a receiver is affected by motion and the presence of gravitational fields. We study the effect of regenerating the signal in intermediate repeaters in different relativistic scenarios and comment the differences with respect to nonrelativistic repeaters.
\end{abstract}
%\pacs{}
\maketitle 

\section{Relativistic noisy channels}
Information Theory studies the limits of communication. In the presence of noise, we can only distinguish reliably a finite number of signals. Shannon's noisy channel theorem establishes a limit to the efficiency of a communication system with signals of average power $P$ subject to additive gaussian noise \cite{Sha49a}. 
We consider signals of length $T$, bandwidth $W$ and average power $P$. These signals can approximate with any desired accuracy real signals, which cannot be strictly limited in both time and frequency \cite{Sle76}. The signal which encodes the data is added during the communication procedure a white noise of power spectral density $N_0$. After filtering the signal at the receiver, we get a total noise power $N=N_0W$. The number of different signals which can be told apart from each other depends on the signal-to-noise ratio, $SNR=P/N$.

The communication efficiency is described in terms of the number of bits that can be sent in a second with negligible error, which is a function of the $SNR$. The maximum possible bit rate is called the channel capacity, which, for the additive gaussian noise channel, is
\begin{equation}
C=W\log_2\left(1+SNR\right).
\end{equation}
This is the asymptotic result of Shannon's theorem, valid when $T\to\infty$. However, it gives a good estimation of the maximum data rate which can be achieved in real channels. 

When relativistic effects are considered, this capacity must be corrected. The signal parameters can be different for different observers. We can describe the situation in the receiver's frame in terms of the signal parameters in the transmitter's frame $T$, $W$ and $P$ and the Doppler factor $\alpha$ which gives the ratio between the frequencies at the receiver $f'$ and the transmitter $f$. 

Data rates and bandwidths are frequencies. In the receiver we have $C'=\alpha C$ and $W'=\alpha W$. Time intervals become divided by $\alpha$ ($T'=T/\alpha$). We assume the receiver can capture the whole signal so that Doppler broadening has no effect on the received power. The signal power at the receiver is then $\alpha^2$ times greater than in the transmitter ($P'=\alpha^2P$). This can be easily understood for an electromagnetic signal. Imagine we have a plane wave of frequency $f$ which is blue-shifted upon reception ($\alpha>1$). The number of photons is conserved, but they have more energy at the receiver ($E'=hf'=h\alpha f=\alpha E$) and arrive in a time window $\alpha$ times smaller. The receiver perceives a power greater by a factor of $\alpha^2$. 

The same factor can be obtained for inertial observers taking the Lorentz transformation of the electric and magnetic fields \cite{LL75,vBla84} or from the constant of motion $Edt=E'dt'$, which is valid for two fixed observers in a stationary gravitational field \cite{TW01,OST00,HSS06,RAO08}. 

If these signals are then added a noise of spectral power density $N_0$ at the receiver, the channel capacity becomes \cite{JC81}
\begin{equation}
C=W\log_2\left( 1+\frac{\alpha P}{N_0W}\right).
\label{relcap}
\end{equation}
The capacity is written in the transmitter's frame (it gives the maximum number of bits the transmitter can send in one second while still expecting the receiver to be able to decode them correctly). The change comes from the $SNR$, which becomes $SNR'=\alpha SNR$. The signal power is multiplied by $\alpha^2$ at the receiver, but the bandwidth is multiplied by $\alpha$. If $\alpha>1$, the signal is stronger, but also occupies a greater bandwidth and we must let more noise through the filter. If $\alpha<1$, the signal is weaker but narrower and we can filter more noise. Nevertheless, if $\alpha>1$ the capacity increases with respect to the nonrelativistic case and when $\alpha<1$ the capacity is smaller (either we need to increase the transmitting power to keep the same data transfer speed or we must slow down the transmission).

\section{Symmetric channels. Inertial users}
We will consider communication situations with two users Alice, $A$, and Bob, $B$, with identical transmitters and receivers. First, we imagine Alice and Bob are in a Minkowski space-time (there are no important masses around them and we can neglect gravitational effects). If Alice and Bob are at relative rest, the channels going from Alice to Bob and from Bob to Alice are identical and their capacity is given by Shannon's formula. If we put the same observers in relative uniform motion, we need to use Equation (\ref{relcap}). We can choose our coordinate system so that Alice is at the origin and Bob is moving at speed $\beta$ in the $x$ direction. All the velocities will be normalized to the speed of light in vacuum so that $-1\le \beta\le 1$. The relativistic Doppler factor for approaching observers is $\alpha=\sqrt{\frac{1+|\beta|}{1-|\beta|}}\ge 1$. For receding observers, $\alpha=\sqrt{\frac{1-|\beta|}{1+|\beta|}}\le 1$. 

The channel capacity for these observers is $C=W\log_2\left(\alpha SNR \right)$. For observers at relative rest ($\alpha=1$), we recover Shannon's formula. Approaching observers can communicate at a faster rate while receding observers must limit the rate of their transmissions. The transmitters and the receivers are in all the cases equal. The change in capacity comes only from the state of motion.

In this case, the channel must be symmetric (the capacity from $A$ to $B$, $C_{AB}$, must be equal to the capacity of the channel from $B$ to $A$, $C_{BA}$). We can always find a frame in which the roles reverse. For a constant $\beta$ both inertial frames can be equally said to be in motion. 

\section{Asymmetric channels}
When we include gravitation or acceleration, there can be asymmetric channels in which $C_{AB}\neq C_{BA}$ \cite{JC81,GC11}. We can see a simple example with the gravitational redshift.  

If Alice is deeper inside a gravitational potential than Bob, her signals are redshifted ($\alpha_{AB}<1$), while Bob's signals are blueshifted ($\alpha_{BA}>1$). In this case, $C_{BA}>C_{AB}$. We will use an example from the redshift in the Schwarzschild metric, but similar results apply for similar situations in other gravitational potentials \cite{OST00,HSS06,RAO08}.

We consider a nonrotating stationary mass $M$. Alice has a fixed station at a distance $r_A$ from the center of mass of $M$ and Bob is at a distance $r_B$. All the distances we consider are above the Schwarzschild radius. Their Doppler factors with respect to an observer far away from the mass ($r\to \infty$) who feels no influence of its gravitational field are (in geometrized units):
\begin{align}
\alpha_{\infty A}=\left(1-\frac{2M}{r_A}\right)^{-\frac{1}{2}} &,& \alpha_{\infty B}=\left(1-\frac{2M}{r_B}\right)^{-\frac{1}{2}}.
\end{align}
Their relative Doppler factor is
\begin{equation}
\alpha_{AB}=\sqrt{\frac{1-\frac{2M}{r_A}}{1-\frac{2M}{r_B}}}.
\label{DopplerSchwarzschild}
\end{equation}

\section{Repeaters in Relativity}
In noisy communication systems with high noise, repeaters can help to overcome the limitations of the channel. We can divide the total path to the receiver into smaller segments in which a better communication is possible. In relativistic channels, we can also use different intermediate frames to improve the end-to-end channel capacity.

We start with a trivial nonrelativistic example. Imagine we have a new observer, Ralph ($R$), between Alice and Bob. $R$ has the same equipment as Alice and Bob: a transmitter with available power $P$ and a receiver with noise $N_0$. We consider a decode-and-forward strategy. Ralph tries to learn the data from Alice's signal and sends it again in a new signal of power $P$ to Bob. In our example where all the noise is added at the receiver, the lossless channel between $A$ and $R$ will be equal to the channel between $R$ and $B$ and both have the same capacity $C=W\log_2(1+\frac{P}{N_0 W})$. 

If we had different noises or powers in each link, we would have two capacities $C_1$ (between $A$ and $R$) and $C_2$ (between $R$ and $B$). The end-to-end capacity $C$ is the minimum of $C_1$ and $C_2$. If $C_1>C_2$, $R$ receives information faster than it can send it. He can store the data in a buffer and send it at a maximum rate $C_2$. If $C_1<C_2$, he could send up to $C_2$ bits per second, but, as he doesn't receive that much information, $R$ must settle for a rate $C_1$. For $C_1=C_2=C$ they have the optimal use of their resources.

In the relativistic case we have a different situation. Consider the channel from Alice to Bob, with a Doppler factor $\alpha$ between Alice's and Bob's frames. Now we have two new channels, one from Alice to Ralph, with a Doppler factor $\alpha_1$, and one from Ralph to Bob, with a factor $\alpha_2$. We have capacities $C_1=W\log_2\left(1+\alpha_1SNR\right)$ and $C_2=W\log_2\left(1+\alpha_2SNR\right)$. Alice can send data with a rate $C_1$, but $R$ will see an arrival rate $\alpha_1C_1$. $R$ can send data at a rate $C_2$.
Table \ref{rates} summarizes the maximum transmission rates and the perceived rate of data arrival at the destination.
 
\begin{table}
 \begin{tabular}{| c | c c c |}
\hline
Channel&Alice&Ralph&Bob\\
\hline
$A\rightarrow B$ & $C$ & & $C'=\alpha C$ \\
\hline
$A\rightarrow R$ & $C_1$ & $C_1'=\alpha_1C_1$&\\
\hline
$R\rightarrow B$ & & $C_2$& $C_2'=\alpha_2C_2$\\
\hline
\end{tabular}
\caption{Maximum rates of communication through a noisy channel. The primed variables correspond to the maximum number of bits that arrive to the receiver per second (as measured in the receiver's frame time).\label{rates}}

\end{table}

If $C_2\ge C_1'=\alpha_1C_1$, we can send the information at the same rate $C_1'$ it arrives. For the repeater's frame best transmission rate, $\alpha_1 C_1$, $B$ sees a rate of arrival $\alpha_1\alpha_2C_1$. In the transmitter's frame, this corresponds to an end-to-end capacity $C=\frac{\alpha_1\alpha_2C_1}{\alpha}$.

If $C_2<C_1'=\alpha_1C_1$, we need to store part of the bits and send the data to Bob at the maximum possible rate $C_2$. $B$ receives bits at a rate $C_2'=\alpha_2C_2$. That corresponds to a capacity of $C=\frac{\alpha_2C_2}{\alpha}$ as seen in the transmitter's frame. 

If we can choose the frame of the repeater, we can choose the best from these capacities. The repeater channel has then a capacity:
\begin{equation}
C=\mathop{\max}_{\alpha_1,\alpha_2}  \min\left(\frac{\alpha_1\alpha_2}{\alpha}C_1, \frac{\alpha_2}{\alpha}C_2\right).
\end{equation}

\begin{figure}[ht!]
\centering
\includegraphics{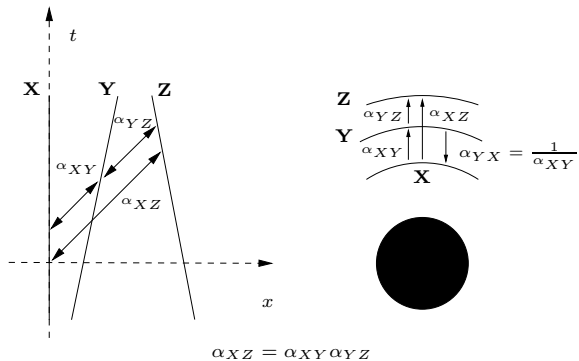}
\caption{\emph{Composition of Doppler factors:} For three observers moving in the $x$ direction or at different depths inside a potential well, we can relate the Doppler factor between the most distant observers with the Doppler factors of both frames with an intermediate point.}
\label{doppcomp}
\end{figure}

We can simplify this formula if we find how $\alpha_1$ and $\alpha_2$ are related to $\alpha$. For observers moving in the $x$ direction (Figure \ref{doppcomp}, left), we can compose Doppler factors by multiplying them. If we have three observers $X$, $Y$ and $Z$ moving in the $x$ direction, we can easily see from the relativistic addition of velocities or an analysis in hyperbolic coordinates that the Doppler factor $\alpha_{XZ}$ for the two most separated observers is the product of the intermediate factors $\alpha_{XY}$ and $\alpha_{YZ}$.

Something similar happens with the gravitational redshift (Figure \ref{doppcomp} right). From the gravitational redshift factors of Equation (\ref{DopplerSchwarzschild}), it is easy to check that
\begin{equation}
\alpha_{XZ}=\sqrt{\frac{1-\frac{2M}{r_X}}{1-\frac{2M}{r_Z}}}=\sqrt{\frac{1-\frac{2M}{r_X}}{1-\frac{2M}{r_Y}}}\sqrt{\frac{1-\frac{2M}{r_Y}}{1-\frac{2M}{r_Z}}}=\alpha_{XY}\alpha_{YZ}.
\end{equation}

In some cases, we can even compose gravitational and motion effects by multiplying their corresponding Doppler factors \cite{RAO08}.

Considering this composition of Doppler factors, there are two interesting situations for a single repeater:

\subsection{$\alpha_1\alpha_2=\alpha$:}
This happens in symmetric channels for repeaters situated in between the two moving end users and in asymmetric channels for any repeater connecting users at different points of a gravitational potential. If we replace $\alpha_2$ by $\frac{\alpha}{\alpha_1}$, we can write everything in terms of $\alpha_1$ and $\alpha$ only. The end-to-end capacity becomes
\begin{equation}
\mathop{\max}_{\alpha_1}\hspace{0.5ex}\min\left(C_1,\frac{C_2}{\alpha_1}\right).
\end{equation}
Capacity $C_1=W\log_2\left(1+\alpha_1SNR\right)$ becomes greater for a greater $\alpha_1$. Capacity $\frac{C_2}{\alpha_1}=\frac{1}{\alpha_1}W\log_2\left(1+\frac{\alpha}{\alpha_1}SNR\right)$ increases with a decreasing $\alpha_1$. The first capacity applies if $C_2\ge \alpha_1 C_1$ and the second if $C_2\le \alpha_1C_1$. The maximum for both (in their respective domains) occurs when $C_2=\alpha_1C_1$. This happens for an $\alpha_1^{*}$ satisfying 
\begin{equation}
W\log_2\left(1+\frac{\alpha}{\alpha_1^{*}}SNR\right)=\alpha_1^{*}W\log_2\left(1+\alpha_1^{*}SNR\right).
\label{alphaopt}
\end{equation}
 
The end-to-end capacity for the optimal repeater frame, $W\log_2\left(1+\alpha_1^{*}SNR \right)$, improves the capacity without a repeater, $W\log_2\left(1+\alpha SNR \right)$, when $\alpha_1^{*}>\alpha$. This condition $1>\frac{\alpha}{\alpha_1^{*}}$ is only satisfied in equation (\ref{alphaopt}) for $1>\alpha_1^{*}>\alpha$. We can only improve the capacity of the channel between receding observers or the uplink communication going out of a gravitational well. 
 
We can see a particular case in the infinite bandwidth approximation ($W\to \infty$), which gives the maximum possible capacity for any given channel with fixed $N_0$ and $P$ values. In the infinite bandwidth limit $C_1=\alpha_1 P/N_0 \log_2(e)$, $C_2=\frac{\alpha}{\alpha_1} P/N_0 \log_2(e)$ and the condition $C_2=\alpha_1C_1$ gives a maximum capacity for $\alpha_{1}^{*}=\sqrt[3]{\alpha}$. The optimal capacity is $C=\sqrt[3]{\alpha}P/N_0\log_2(e)$. In this example, we can easily check there is only a gain with respect to the case without repeaters, with $C=\alpha P/N_0\log_2(e)$, if $\alpha<1$ (the receiver is receding). 

\begin{figure}[ht!]
\centering
\includegraphics{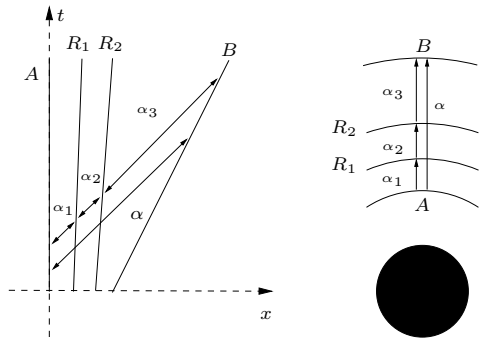}
\caption{\emph{Chain of repeaters:} We can introduce multiple repeaters between Alice and Bob. We give two examples for inertial observers moving away from Alice (left) and repeaters at different points of a gravitational potential (right).}
\label{redchan}
\end{figure}

We can also extend the argument to multiple repeaters. Imagine Alice and Bob are at fixed stations over a nonrotating planet of mass $M$ at distances $r_A$ and $r_B$ respectively, with $r_B>r_A$. We will consider a series of repeaters at different heights over the surface of the planet (Figure \ref{redchan}, right). 

For a chain of $n$ repeaters, we must choose the $n$ frames which optimize the total capacity. We can divide the channel in two parts, a first channel from Alice to the last repeater and a channel between the last repeater and the destination. Our capacities are given in the frame of Alice. If all the previous repeaters give a total capacity $C_n$, the incoming data rate in repeater $n$ is $\alpha_nC_n$. The situation is equivalent to the case with just one repeater and the condition $\alpha_nC_n=C_{n+1}$ gives the best final capacity. The end-to-end capacity in that case is $\frac{\alpha_n\alpha_{n+1}C_n}{\alpha}$. We can maximize this capacity if we optimize $C_n$.

\begin{figure}[ht!]
\centering
\includegraphics{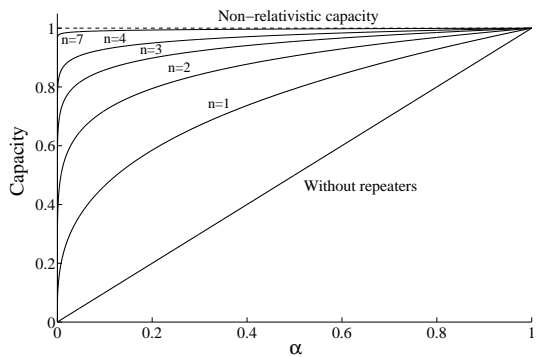}
\caption{\emph{Effect of repeaters on a redshifted channel:} We show the normalized capacity for an infinite bandwidth channel ($P/N_0\log_2 e=1$) between a transmitter deeper inside a potential well than the receiver. The curves represent the normalized capacity for different redshift factors $\alpha$ and a different number of repeaters $n$. With a small number of repeaters we can come close to the capacity the same two users would achieve outside the gravitational potential.}
\label{chainrep}
\end{figure}

We can imagine now repeater $n$ is the final destination and apply the same argument for the maximization condition at repeater $n-1$. We can repeat this procedure as many times as necessary. The end-to-end capacity will be maximal if 
\begin{equation}
\alpha_iC_i=C_{i+1}
\label{reccond}
\end{equation} 
for all repeaters (from $i=1$ to $n$), which gives a capacity
\begin{equation}
\frac{\alpha_1\alpha_2\ldots\alpha_{n+1}}{\alpha}C_1=C_1.
\end{equation}
The last step comes from noticing that the product of all the factors must give the total $\alpha$ between Alice and Bob. For a high number of repeaters, we can improve the value of $\alpha_1$ and, therefore, of $C_1$. We can see an example in the infinite bandwidth approximation. Condition (\ref{reccond}) becomes
\begin{equation}
\alpha_i^2 P/N_0 \log_2 e=\alpha_{i+1} P/N_0 \log_2 e. 
\end{equation}
The Doppler factor must be squared at each step ($\alpha_2=\alpha_1^2$, $\alpha_3=\alpha_2^2=\alpha_1^4$ \ldots). We can find the optimal $\alpha_1^{*}$ which determines all the $\alpha_i$ from
\begin{equation}
\alpha=\prod_{j=1}^{n+1}\alpha_j=\prod_{k=0}^{n}{\alpha_1^{*}}^{2^k}={\alpha_1^{*}}^{\sum_{k=0}^{n}2^k}={\alpha_1^{*}}^{2^{n+1}-1},
\end{equation}
with an optimal $\alpha_1^{*}=\alpha^{\frac{1}{2^{n+1}-1}}$. As in the single repeater case, we can only improve the capacity if $\alpha<1$. Figure \ref{chainrep} shows an example on the effect of adding repeaters for different redshift factors $\alpha$.

We can also consider a chain of repeaters between inertial observers in relative motion. Imagine a rocket going to Proxima Centauri at high speed. A control space station in the Solar System will see an $\alpha<1$. The capacity will be smaller than in the nonrelativistic case. We can imagine a float of ``Little Thumb'' repeaters: the rocket can drop a repeater at each acceleration stage. Each of the repeaters will be moving at a smaller speed than the original ship (see Figure \ref{redchan}, left) and the analysis is totally equivalent to the gravitational well example. With a few repeaters we can greatly improve the capacity until we are close to the limiting case with no relativistic effects. However, in the next section we show we can improve this bound in a more general scenario.

\subsection{$\frac{\alpha_1}{\alpha_2}=\alpha$:}
If the destination is between the transmitter and the repeater instead of having a repeater in the middle, the relationship between observers is different. Now, the Doppler factor between the most distant extremes (the channel from Alice to the repeater) is $\alpha_1$ and $\alpha_1=\alpha\alpha_{2}$.

\begin{figure}[ht!]
\centering
\includegraphics{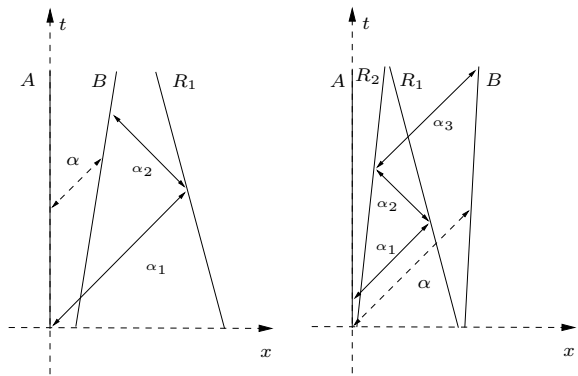}
\caption{\emph{Arbitrary capacity for approaching repeaters:} If we have one repeater approaching both Alice and Bob, we can obtain an end-to-end capacity as high as desired (left). Two repeaters between Alice and Bob can also produce the same effect (right). In both cases, the total delay between transmission and reception increases and we need a memory buffer (larger as $C$ increases).}
\label{infinite}
\end{figure}

The channel capacity becomes
\begin{equation}
\mathop{\max}_{\alpha_1}\hspace{0.5ex}\min\left(\frac{\alpha_1^2}{\alpha^2} C_1,\frac{\alpha_1 C_2}{\alpha^2}\right).
\end{equation}
In this situation, we can place ourselves in the second regime (where $\alpha_1C_1>C_2$ and we need a buffer) and obtain an arbitrarily high capacity. For any fixed $\alpha$, greater or smaller than $1$, we can always find an $\alpha_1$ high enough to make capacity $\alpha_1C_1=\alpha_1W\log_2\left(1+\alpha_1 SNR\right)$ greater than $C_2=W \log_2\left(1+\frac{\alpha_1}{\alpha}SNR\right)$. The end-to-end capacity will be $\frac{\alpha_1}{\alpha^2}W \log_2\left(1+\frac{\alpha_1}{\alpha}SNR\right)$, which can be made as large as desired if we have a repeater approaching the sender close to the speed of light. We can always choose a frame approaching both Alice and Bob with high capacity in both channels (see Figure \ref{infinite}, left). The price to pay is a buffer (which will have to be larger for higher capacity values).

A similar analysis shows that, even if all the repeaters are between Alice and Bob, two repeaters are enough to guarantee as high a capacity as desired (Figure \ref{infinite}, right). If we choose the second repeater $R_2$ to be approaching Bob (going away from Alice at a higher speed) and the first repeater $R_1$ to be approaching Alice and $R_2$, all the channels can have a higher capacity than the nonrelativistic channel.

For these four observers $\alpha=\frac{\alpha_1\alpha_3}{\alpha_2}$. We can make the Doppler factors between Alice and $R_1$ ($\alpha_1$) and between $R_2$ and Bob ($\alpha_3$) as high as desired. For high values of these factors, we can guarantee that in each repeater the data arrives at a faster rate than it can be sent. In this buffered regime, we can obtain an end-to-end capacity $\frac{\alpha_3}{\alpha}W\log_2\left(1+\alpha_3 SNR\right)$, which can be made as high as desired.

For approaching repeaters, we can obtain, in principle, an infinite capacity. However, there are a few important remarks. Capacity is measured as the maximum sustained rate of information transmission which is possible. Delays, such as those that happen in decoding, are irrelevant. Here, we have an additional source of delay, which is the time it takes the data to zigzag through all the repeaters. Additionally, if capacity tends to infinity, we also need an infinite buffer at the repeaters. 

Taking these factors into account, the scenarios of the previous section where $\alpha_1\alpha_2=\alpha$ seem more natural. The space journey example, for instance, is close to what we can expect for a real interplanetary mission, where distant repeaters have not been deployed and we have a limited speed (below the speed of light). 

\section{Discussion} 

We have seen the differences between relativistic and nonrelativistic communication. The maximum communication rate between two observers depends on their state of motion and whether there is a gravitational field or not. Communication between observers going away from each other or climbing up a gravitational potential is less efficient than the communication between observers with no relative motion and far from any mass. A transmission between approaching observers or a communication going down a gravitational potential can be done at a higher rate. 

Introducing intermediate nodes with a decode-and-forward strategy has no effect in the nonrelativistic capacity. In relativistic channels where only the receiver adds noise, we can improve the less efficient cases and, in some cases, we can even attain arbitrarily high capacities. 

There are many interesting generalizations to the repeater channel we have introduced. Our starting point has been a nonrelativistic decode-and-forward strategy. The decode-and-forward channel is a particular, trivial case of the relay channel \cite{CG79}. A relativistic generalization of the full relay channel could show new unexpected properties of relativistic communications.

The simple examples we have presented in this paper show that Relativity can help us to understand better the physical limits of communication and allows us to devise improved communication protocols. 

\section*{Acknowledgments}
This work has been funded by projects Junta de Castilla y Le\'on VA342B11-2 and MICINN TEC2010-21303-C04-04.

\newcommand{\noopsort}[1]{} \newcommand{\printfirst}[2]{#1}
  \newcommand{\singleletter}[1]{#1} \newcommand{\switchargs}[2]{#2#1}


\begin{thebibliography}{10}

\bibitem{Sha49a}
C.E. Shannon.
\newblock Communication in the presence of noise.
\newblock {\em Proceedings of the IRE}, 37(1):10--21, 1949.

\bibitem{Sle76}
D.~Slepian.
\newblock On bandwidth.
\newblock {\em Proceedings of the IEEE}, 64(3):292 -- 300, 1976.

\bibitem{LL75}
L.D. Landau and E.M. Lifshitz.
\newblock {\em The Classical Theory of Fields}, volume~2 of {\em Course of
  Theoretical Physics}.
\newblock Pergamon Press, fourth edition, 1975.

\bibitem{vBla84}
J.~{van Bladel}.
\newblock {\em Relativity and Engineering}.
\newblock Springer Series in Electrophysics, Volume 15. Springer-Verlag,
  Berlin, 1984.

\bibitem{TW01}
Edwin Taylor and John Wheeler.
\newblock {\em Exploring Black Holes: Introduction to General Relativity}.
\newblock Addison-Wesley, Boston, 2001.

\bibitem{OST00}
L.~B. Okun, K.~G. Selivanov, and V.~L. Telegdi.
\newblock On the interpretation of the redshift in a static gravitational
  field.
\newblock {\em American Journal of Physics}, 68(2):115--119, 2000.

\bibitem{HSS06}
Alex Harvey, Engelbert Schucking, and Eugene~J. Surowitz.
\newblock Redshifts and Killing vectors.
\newblock {\em American Journal of Physics}, 74(11):1017--1024, 2006.

\bibitem{RAO08}
A.~Radosz, AT~Augousti, and K.~Ostasiewicz.
\newblock {Decoupling of kinematical time dilation and gravitational time
  dilation in particular geometries}.
\newblock {\em Acta Physica Polonica B}, 39(6):1357, 2008.

\bibitem{JC81}
Keith Jarett and Thomas~M. Cover.
\newblock Asymmetries in relativistic information flow.
\newblock {\em IEEE Transactions on Information Theory}, 27(2):152--159, 1981.

\bibitem{GC11}
J.~C. {Garcia-Escartin} and P.~{Chamorro-Posada}.
\newblock {Secure communication in the twin paradox}.
\newblock {\em arXiv:1006.2705v2}, 2011.

\bibitem{CG79}
T.~Cover and A.E. Gamal.
\newblock Capacity theorems for the relay channel.
\newblock {\em IEEE Transactions on Information Theory}, 25(5):572 -- 584, sep
  1979.

\end{thebibliography}
\end{document}